\documentclass[epj,twocolumn, raggedbottom]{svjour}
\usepackage{latexsym}
\usepackage{graphics}
\usepackage{graphicx}
\usepackage{amsmath}
\usepackage{color}
\usepackage{braket}
\usepackage{float}

\begin{document}
\title{Stereochemical properties of the OH molecule in combined electric and magnetic fields: analytic results}
\author{Stefano Marin and Mishkatul Bhattacharya
}                     
\institute{School of Physics and Astronomy, Rochester Institute of Technology,
84 Lomb Memorial Drive, Rochester, NY 14623, USA}
%
\date{\today}
%
\abstract{
The stereochemical properties of the ultracold ground state OH molecule in the presence of electric and
magnetic fields are currently of considerable interest. For example, relevant quantities such as molecular
alignment and orientation, calculated numerically by using large basis sets, have lately appeared in the
literature. In this work, based on our recent exact solution to an effective eight-dimensional matrix
Hamiltonian for the molecular ground state, we present \textit{analytic} expressions for the stereochemical
properties of OH. Our results require the solution of algebraic equations only, agree well with the
aforementioned fully numerical calculations, provide compact expressions for simple field geometries,
allow ready access to relatively unexplored parameter space, and yield straightforwardly higher moments
of the molecular axis distribution.
%
} 
\titlerunning{Stereochemical properties of the OH molecule}
\maketitle

\section{Introduction}
\label{sec:Intro}
The ultracold ground state $X^{2}\Pi_{3/2}$ OH molecule is a versatile enabler for studies of
spectroscopy \cite{Hudson2006}, precision measurement \cite{Lev2006,Kozlov2009,Alyabyshev2012},
trapping \cite{Sawyer2007}, slowing \cite{Bochinski2004,Meerakker2005}, state transfer \cite{Stuhl2012},
cold chemistry \cite{Avdeenkov2003,Ticknor2005,Sawyer2008,Tscherbul2010,Quemener2013}, and quantum
degeneracy \cite{Quemener2012}. The molecular ground state is polar as well as paramagnetic, and
thus many theoretical \cite{Lara2008,Bohn2013,MBEPJD2013,Mishkat2013,Mishkat2014,Maeda2015} and experimental
\cite{Lemeshko2013} studies have focused on the behavior of ultracold OH in combined electric and
magnetic fields.

Recently, the stereochemical properties of cold molecules have come under investigation, due to
their importance for cold chemistry \cite{Miranda2011,Stuhl2014,Sharma2015} and molecular beam
manipulation \cite{Friedrich2006}, for example. Particularly relevant is the work by
G\"{a}rttner et al. \cite{Schmelcher2013}, who have calculated the OH molecular axis alignment and
orientation by numerically solving the time-independent Schrodinger equation using large sets of
basis functions. While such numerical investigations lead to accurate and comprehensive characterization
of the OH properties, analytical studies have also provided a fruitful line of investigation in the past,
yielding physical insight as well as mathematical compactness \cite{Lara2008,Bohn2013,MBEPJD2013,MB2014}.
In particular, it has been found that experimental data on Landau-Zener transitions \cite{Stuhl2012} as
well as evaporative cooling \cite{Quemener2012} can be accurately modeled using an eight-level
characterization of the ground state OH molecule in crossed electric and magnetic fields \cite{Lara2008};
we have recently analytically diagonalized the corresponding Stark-Zeeman matrix Hamiltonian with the
assistance of an underlying chiral symmetry \cite{Mishkat2013,Mishkat2014}.

In this article, we exploit the availability of this solution to calculate analytical expressions for the
alignment and orientation of the ground state OH molecule. We find good agreement between our analytical
results and the existing numerical calculations, indicating the reliability of our expressions. We present
our results in order of increasing complexity, by first considering simple field geometries. These yield
quite compact expressions - which are exact within the framework of the effective matrix Hamiltonian - for the
stereochemical properties. Subsequently, we consider more complicated configurations, where some approximate
but useful expressions are presented. Then we show that our full expressions can be used to directly
investigate relatively unexplored parameter regimes, such as variation in the angle between the two fields.
Finally, we demonstrate that higher order moments of the molecular axis distribution \cite{Hockett2015} can
also be found simply using our approach.

The remainder of this paper is arranged as follows. Section~\ref{sec:Form} establishes the formalism using
the effective Hamiltonian, Section ~\ref{sec:Stereo} discusses the stereochemical properties, and Section
~\ref{sec:Con} supplies the conclusions.

\section{Formalism}
\label{sec:Form}
\begin{figure}[h]
\centering
\includegraphics[width=0.45\textwidth]{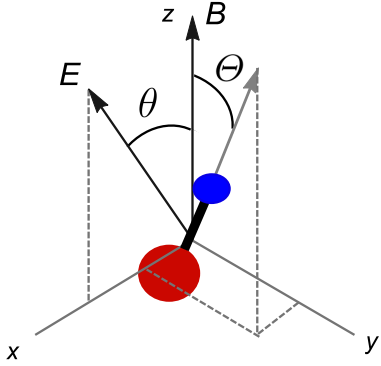}
\caption{Schematic of the diatomic OH molecule in external magnetic $(B)$ and electric $(E)$ fields. The magnetic field
has been chosen to lie along the laboratory $z$ axis and the electric field on the $x-z$ plane, making an angle
$\theta$ with the $z$ axis. The angle between the laboratory $z$ and molecular axes is labelled $\Theta$ and
is the main interest of this paper.}
\label{fig:F1}
\end{figure}

In this section we outline the formalism used to calculate the OH stereochemical properties. We first
recapitulate the ground state matrix OH Hamiltonian derived originally by Lara et al. \cite{Lara2008},
for the system shown in Fig.~\ref{fig:F1}.
The Hamiltonian neglects hyperfine structure and spin-orbit interactions, is valid for strong, but not
ultrastrong electric and magnetic fields, and accounts for the Lambda-doublet in the ground OH rotational
state
\begin{equation}
\label{eq:H1}
H=H_{o}-\vec{\mu}_{e}\cdot \vec{E}-\vec{\mu}_{b}\cdot \vec{B}.
\end{equation}
In Equation ~(\ref{eq:H1}), $H_{o}$ is the field-free Hamiltonian, the second term accounts for the Stark
interaction and the third term corresponds to the Zeeman interaction. The symbol $\vec{\mu}_{e}(\vec{\mu}_{b})$
stands for the molecular electric (magnetic) dipole moment, and $\vec{E}(\vec{B})$ is the static electric
(magnetic) field. To represent this Hamiltonian in matrix form, we use the Hund's case (a) parity basis
$|J, M, \bar{\Omega}, \epsilon \rangle$ following Lara et al. \cite{Lara2008}, where $J = 3/2$ is
the rotational angular momentum, $M$ is its projection in the laboratory frame, $\bar{\Omega}$ is its projection on
the internuclear axis, and $\epsilon = \{e,f \}$ is the $e-f$ symmetry. In this representation, the Hamiltonian
reads \cite{MBEPJD2013}


\begin{equation}
\label{eq:Hmatrix}
H_M=
\left(
\begin{array}{cc}
A_{1}-A_{2} & -C\\
-C & A_{1}+A_{2}\\
\end{array} \right),
\end{equation}
with
\begin{align}
A_{1} =\frac{2}{5} \mu_B B
&\left( \begin{array}{cccc}
-3 & 0 & 0 & 0\\
0 & -1 & 0 & 0 \\
0 & 0 & 1 & 0 \\
0 & 0 & 0 & 3\\
\end{array} \right), \\
A_{2} = \frac{\hbar \Delta}{2} &\left( \begin{array}{cccc}
1 & 0 & 0 & 0\\
0 & 1 & 0 & 0 \\
0 & 0 & 1 & 0\\
0 & 0 & 0 & 1\\
\end{array} \right), \\
C = \frac{\mu_{e} E}{5} &\left( \begin{array}{cccc}
-3 \cos \theta &  \sqrt{3} \sin \theta & 0 & 0 \\
\sqrt{3} \sin \theta & -\cos \theta & 2 \sin \theta & 0 \\
0 & 2 \sin \theta &  \cos \theta & \sqrt{3} \sin \theta \\
0 & 0 &  \sqrt{3} \sin \theta & 3 \cos \theta \\
\end{array} \right),\\
\nonumber
\end{align}
where the Lambda-doubling parameter is denoted by $\Delta$, the Bohr magneton by $\mu_{B}$, the magnitude
of the molecular electric dipole moment by $\mu_e$, the electric and magnetic field magnitudes by
$E=|\vec{E}|$ and $B=|\vec{B}|$, respectively, and the angle between the magnetic and electric field
vectors by $\theta$. For convenience, below we will use the quantities
\begin{equation}
\tilde{E} = \mu_{e} E, \, \tilde{B} = \mu_{B} B, \, \tilde{\Delta} = \hbar \Delta,
\end{equation}
all of which have dimensions of energy. The numerical values of the relevant constants are $\mu_{e}=1.66$D,
$\mu_{B}=9.27\times 10^{-24}$JT$^{-1}$ and $\Delta=2\pi\times 1.67$GHz.

The eigenvalues of the eight-dimensional matrix $H_{M}$ of Eq.~(\ref{eq:Hmatrix}) can be found analytically,
as shown earlier \cite{Mishkat2013}. For the calculations in this article, we used the commercial software
package \textit{Mathematica} to find the eigenstates of $H_{M}$ analytically. The typical form of the OH
eigenstate is
\begin{equation}
\label{eq:eigenv}
\ket{\psi}=\frac{1}{N}\ket{c_{1},c_{2},c_{3},c_{4},c_{5},c_{6},c_{7},c_{8}},
\end{equation}
where the $c_{i}$ are the components in the Hund's case (a) parity basis, and the normalization is given by
\begin{equation}
N=\sqrt{\sum_{i=1}^{8}c_{i}^{2}}.
\end{equation}
The $c_{i}$'s, which are always real, and are somewhat large in form, have been provided in Appendix A.

\section{Stereochemical properties}
\label{sec:Stereo}
In this section we present our analytical results for the OH stereochemical properties.
We consider the expressions for quantities of the form $\langle \cos^{k}\Theta\rangle$,
which include the orientation $(k=1)$, alignment $(k=2)$ as well as higher moments
$(k=3,4)$ of the molecular axis distribution \cite{Hockett2015}. The angle $\Theta$ refers
to the inclination of the molecular axis with respect to the laboratory $z$ axis, see Fig.1.
Using the matrix elements of $\cos \Theta$ in the Hund's case (a) basis from Sharma et al.
\cite{Sharma2015}, we find, for the eigenstate of Eq.~(\ref{eq:eigenv}),
\begin{equation}
\label{eq:Cos1}
\langle \cos{\Theta} \rangle = - \frac{2}{5 N^2} \left( 3 c_1 c_5 + c_2 c_6 -c_3 c_7 -3 c_4 c_8 \right ),
\end{equation}

\begin{equation}
\begin{aligned}
\label{eq:Cos2}
\langle \cos^2{\Theta} \rangle =  \frac{1}{15 N^2 }& \left[ 7 \left(c_1^2 + c_4^2 + c_5^2+ c_8^2\right) \right. \\
 &\left.+ 3\left( c_2^2 + c_3^2 + c_6^2 + c_7^2 \right) \right],
\end{aligned}
\end{equation}

\begin{equation}
\label{eq:Cos3}
\langle \cos^3{\Theta} \rangle = - \frac{2}{25 N^2} \left( 7 c_1 c_5 + c_2 c_6 -c_3 c_7 -7 c_4 c_8 \right ),
\end{equation}

\begin{equation}
\begin{aligned}
\label{eq:Cos4}
\langle \cos^4{\Theta} \rangle =  \frac{1}{1225 N^2 }& \left[ 289 \left(c_1^2 + c_4^2 + c_5^2+ c_8^2\right) \right. \\
 &\left.+ 89\left( c_2^2 + c_3^2 + c_6^2 + c_7^2 \right) \right].
\end{aligned}
\end{equation}
In the interest of readability, we start by considering simple geometries for which we find compact analytic
expressions, and then move on to more general cases. Readers who wish to first convince themselves that our
analytic expressions match well with previous numerical calculations may refer to Section~\ref{subsec:Crossed},
especially Fig.~\ref{fig:F6}, directly.

\subsection{Magnetic Field Only}
\label{subsec:BField}


We start by considering the orientation and alignment in the presence of a magnetic field only,

\begin{equation}
\langle \cos{\Theta} \rangle = 0, \, \,  \text{for all states,}
\end{equation}

\begin{equation}
\langle \cos^2{\Theta} \rangle = \frac{1}{30} \left(4 M^2+5\right).
\end{equation}

Thus, all states with $M = \pm 3/2$ have $\langle \cos^2{\Theta} \rangle = 7/15$, whereas for $M = \pm 1/2$,
$\langle \cos^2{\Theta} \rangle = 1/5$. We see that the magnetic field does not affect either of these
stereochemical properties; this is because the Zeeman terms in Eq.~(\ref{eq:H1}) are diagonal in the Hund's
case (a) basis, and therefore do not influence the components of the eigenstate [Eq.~(\ref{eq:eigenv})] at
all. The presence of a magnetic field is thus equivalent to the field-free case.

\subsection{Electric field only}
\label{subsec:EField}
\begin{figure}[h]
\centering
\includegraphics[width=0.45\textwidth]{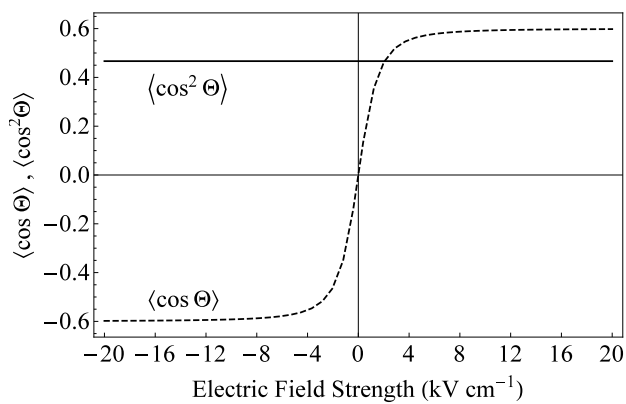}
\caption{$\langle \cos{\Theta} \rangle$ (dashed), $\langle \cos^2{\Theta} \rangle$ (solid) for $B = 0$ as a function of
the electric field.}
\label{fig:F2}
\end{figure}

We now consider the orientation and alignment in the presence of an electric field only. In this case, it
is convenient to rotate the electric field to the laboratory $z$ axis in Fig. ~\ref{fig:F1}. While general
expressions applying to all states can easily be given, for the sake of illustration, we discuss the
properties of the lowest energy state $\ket{\psi_{g}}$. In the presence of an electric field this state is
given by
\begin{align}
\ket{\psi_{g}} = \frac{1}{N} \ket{0,0,-\frac{-5 \tilde{\Delta} -\sqrt{25 \tilde{\Delta} ^2+4  \tilde{E}^2}}{2  \tilde{E}},0,0,0,1,0},
\end{align}
in the Hund's case (a) parity basis with
\begin{equation}
N = \sqrt{2+\frac{5 \tilde{\Delta}  \left(5 \tilde{\Delta} +\sqrt{25 \tilde{\Delta} ^2+36 \tilde{E}^2}\right)}{18 \tilde{E}^2}}.
\end{equation}
A straightforward calculation yields the orientation
\begin{equation}
\langle \cos{\Theta} \rangle = \displaystyle \frac{18 \tilde{E} }{5 \sqrt{25 \tilde{\Delta} ^2+36 \tilde{E} ^2}},
\end{equation}
and the alignment
\begin{equation}
\langle \cos^2{\Theta} \rangle = \displaystyle \frac{7}{15}.
\end{equation}
The alignment is independent of the electric field due to the presence of chiral symmetry \cite{Mishkat2014}.
The orientation, as can be seen from the property
\begin{equation}
\frac{\partial \langle \cos{\Theta} \rangle }{\partial \tilde{E}}=\frac{90\tilde{\Delta}^{2}}{\left(25 \tilde{\Delta}^{2}+36 \tilde{E}^{2}\right)^{3/2}},
\end{equation}
increases monotonically with the electric field, and reaches an asymptotic value of
\begin{equation}
\lim_{\tilde{E}\rightarrow \infty}\langle \cos{\Theta} \rangle=\frac{3}{5}.
\end{equation}
The orientation and alignment have been plotted versus the electric field in Fig.~\ref{fig:F2}.

\subsection{Parallel fields}
\label{subsec:Parallel}
As we have shown in Section~\ref{subsec:BField}, a magnetic field along the laboratory $z$ axis does not
influence the orientation or alignment. If, in addition, an electric field is introduced the same direction
(i.e. $\theta=0$ in Fig.~\ref{fig:F1}), then for these parallel fields the results of Section~\ref{subsec:EField}
apply. In other words, only the electric field has an effect on the stereochemical properties of the molecule
and it influences only the molecular orientation.

\subsection{Perpendicular fields}
\label{subsec:Perp}
We now consider the case where the electric and magnetic fields are applied at right angles to each other,
i.e. $\theta=\pi/2$ in Fig.~\ref{fig:F1}. We can express the lowest energy eigenstate as

\begin{small}
\begin{equation}
\begin{aligned}
\ket{\psi_{g}}=\frac{1}{N} & |0,\frac{\left(\gamma-12\tilde{B}+5\tilde{\Delta}\right)
\left[2\tilde{B}\left(\gamma+8\tilde{B}\right)-2\omega+\tilde{E}^{2}\right]}
{2\tilde{E}\left[4\tilde{B}\left(5\tilde{\Delta}-4\tilde{B}\right)-4\omega+5\tilde{E}^{2}\right]},0,\\
&\frac{2 \sqrt{3} \tilde{E}}{\gamma+12 \tilde{B}-5 \tilde{\Delta} },\frac{\sqrt{3}\tilde{E} ^2}
{-2 \gamma \tilde{B}+16 \tilde{B}^2-2 \omega+\tilde{E} ^2},0,1,0\rangle,\\
\end{aligned}
\end{equation}
with
\begin{equation}
\begin{aligned}
N &  = \left(\frac{3 \tilde{E}^4}{\left(16 \tilde{B}^2-2 \tilde{B} \gamma +\tilde{E}^2-2 \omega \right)^2}+\frac{12 \tilde{E}^2}{(12 \tilde{B}+\gamma -5 \tilde{\Delta} )^2}\right.\\
&\left. \frac{(-12 \tilde{B}+\gamma +5 \tilde{\Delta} )^2 \left[2 \tilde{B} (8 \tilde{B}+\gamma )+\tilde{E}^2-2 \omega \right]^2}{4 \tilde{E}^2 \left(4 \tilde{B} (5 \tilde{\Delta} -4 \tilde{B})+5 \tilde{E}^2-4 \omega \right)^2}+1 \right)^{\frac{1}{2}},
\end{aligned}
\end{equation}
\end{small}
where

\begin{small}
\begin{equation}
\omega = -\sqrt{25 \tilde{B}^2 \tilde{\Delta} ^2+5 \tilde{B} \tilde{\Delta}  \left(8 \tilde{B}^2-\tilde{E}^2\right)+\left(4 \tilde{B}^2+\tilde{E}^2\right)^2},
\end{equation}
\end{small}
and

\begin{figure}[h]
\centering
\includegraphics[width=0.45\textwidth]{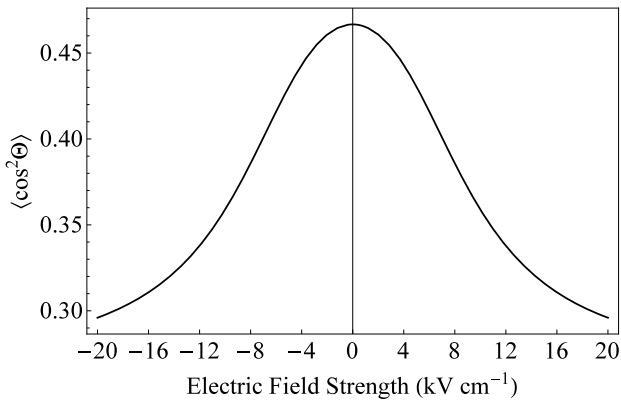}
\caption{$\langle \cos^2{\Theta} \rangle$ versus electric field for $\theta = \pi/3$ and $B = 0.2T$.}
\label{fig:F3}
\end{figure}

\begin{figure}[h]
\centering
\includegraphics[width=0.45\textwidth]{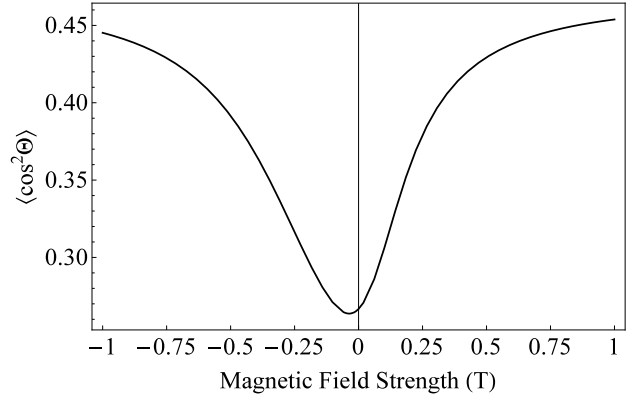}
\caption{$\langle \cos^2{\Theta} \rangle$ versus magnetic field for $\theta = \pi/2$ and $E = 10 \text{k}V \text{cm}^{-1}$.}
\label{fig:F4}
\end{figure}

\begin{figure}[h]
\centering
\includegraphics[width=0.45\textwidth]{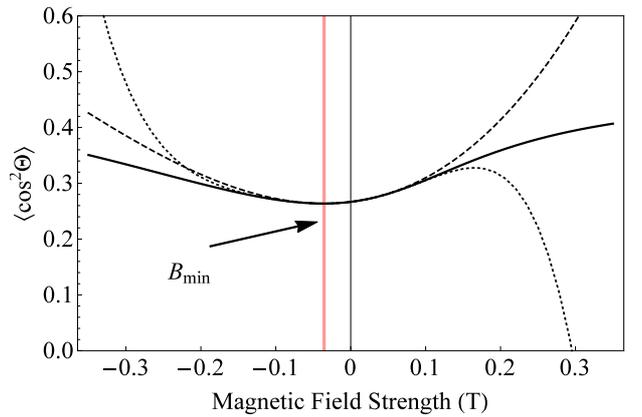}
\caption{$\langle \cos^2{\Theta} \rangle$ from Eq.~(\ref{eq:AlPerp})(solid line), approximations from
Eq.~(\ref{eq:Taylor1}) to 3rd order (dashed line) and 5th order (dot-dashed line). The vertical line
indicates the value of the alignment minimum from Eq.~(\ref{eq:Bmin}). The parameters are
$\theta = \pi/2$ and $E = 10 \text{k}V \text{cm}^{-1}$.}
\label{fig:F5}
\end{figure}

\begin{small}
\begin{equation}
\gamma = \sqrt{80 \tilde{B}^2+40 \tilde{B} \tilde{\Delta} +25 \tilde{\Delta} ^2+20 \tilde{E}^2 - 16\omega}.
\end{equation}
\end{small}
For the orientation of the state $\ket{\psi_{g}}$, we find
\begin{equation}
\label{eq:CosPerp}
\langle \cos{\Theta} \rangle = 0,
\end{equation}
which turns out to be true for all eight states (see also Fig.~\ref{fig:F7} below). For the alignment, we get

\begin{equation}
\label{eq:AlPerp}
\langle \cos^2{\Theta}\rangle =  \displaystyle \frac{\frac{\nu}{540 \tilde{E}^6}+\frac{(12 \tilde{B}+\gamma +5 \Delta )^2}{60 \tilde{E}^2}+\frac{\left[2 \tilde{B} (8 \tilde{B}+\gamma )+\tilde{E}^2-2 \omega \right]^2}{15 \tilde{E}^4}+\frac{7}{15}}{\frac{\nu}{36 \tilde{E}^6}+\frac{(12 \tilde{B}+\gamma +5 \tilde{\Delta} )^2}{12 \tilde{E}^2}+\frac{\left[2 \tilde{B} (8 \tilde{B}+\gamma )+\tilde{E}^2-2 \omega \right]^2}{3 \tilde{E}^4}+1},
\end{equation}
where
\begin{equation}
\begin{aligned}
\nu =&\left\{96 \tilde{B}^3+8 \tilde{B}^2 (\gamma +20 \tilde{\Delta} )\right.\nonumber \\
&+2 \tilde{B} \left[5 \tilde{\Delta}  (\gamma +5 \Delta )+6 \tilde{E}^2-12 \omega \right] \nonumber\\
&\left. -(\gamma +5 \tilde{\Delta} ) \left(\tilde{E}^2+2 \omega \right)\right\}^2,
\end{aligned}
\end{equation}
The alignment has been plotted in Fig.~\ref{fig:F3}
as a function of the electric field, for specific values of $B$ and $\theta$ (please see the figure
caption for values). As expected from Eq.~(\ref{eq:AlPerp}), the alignment is even in $\tilde{E}$.
The alignment is plotted in Fig.~\ref{fig:F4} as a function of the magnetic field
and is not even in $\tilde{B}$. In order to describe Fig.~\ref{fig:F4} in useful detail, we consider a
simpler, but approximate, characterization of $\langle \cos^2{\Theta} \rangle$ by taking a Taylor
series expansion of Eq.~(\ref{eq:AlPerp}) around $B = 0$
\begin{small}
\begin{align}
\label{eq:Taylor1}
\langle \cos^2{\Theta} \rangle = & \frac{4}{15}+\frac{\tilde{B} \tilde{\Delta}}{2 \tilde{E}^2}
+\frac{\tilde{B}^2 \left(75 \tilde{\Delta} ^2+32 \tilde{E}^2\right)}{40 \tilde{E}^4}\\ \nonumber
&+\left(\frac{\tilde{B}}{\tilde{E}}\right)^3 \left(\frac{25 \tilde{\Delta} ^3}{16 \tilde{E}^3}+\frac{2 \tilde{\Delta} }{\tilde{E}}\right)
+ \mathcal{O}(\tilde{B}^4).
\end{align}
\end{small}
From Fig.~\ref{fig:F5} it can be seen that using the Taylor series of Eq.~(\ref{eq:Taylor1}) and retaining
the term up to $\tilde{B}^3$ (dashed line) describes quite well the $\langle \cos^2{\Theta} \rangle$, in
the magnetic field interval $ ~[-0.1, 0.1] T$. For comparison, a Taylor approximation up to fifth order has
also been shown using a dot-dashed line.
The approximate location of the alignment minimum can be found from Eq.~(\ref{eq:Taylor1})
\begin{equation}
\label{eq:Bmin}
B_{\mathrm{min}} \approx \frac{2 \left(-32 \tilde{E}^4-75 \Delta ^2 \tilde{E}^2+\sqrt{1024 \tilde{E}^8
+1875 \tilde{\Delta} ^4 \tilde{E}^4}\right)}{15 \tilde{\Delta}\left(25 \tilde{\Delta}^2+32\tilde{E}^2\right)}.
\end{equation}
This expression takes the value $B_{\mathrm{min}} \approx -0.0355 T$ for $\theta = \pi/2$ and
$E = 10 \text{k}V \text{cm}^{-1}$, and, as can be seen from the vertical line in Fig.~\ref{fig:F5}, is
quite a good approximation.
\subsection{Crossed fields}
\label{subsec:Crossed}
In this section we consider the most general case, with both fields turned on, and a variable angle of
separation between them. Fully analytic expressions for the results were found by combining the formulas
of Eqs.~(\ref{eq:Cos1}) and (\ref{eq:Cos2}) with those of Appendix A, but are quite long, and will not be
presented them here. Instead the intention behind this section is to show that these analytic formulas are
very convenient for plotting quantities of interest versus experimentally tunable parameters.

\begin{figure}[h]
\centering
\includegraphics[width=0.5\textwidth]{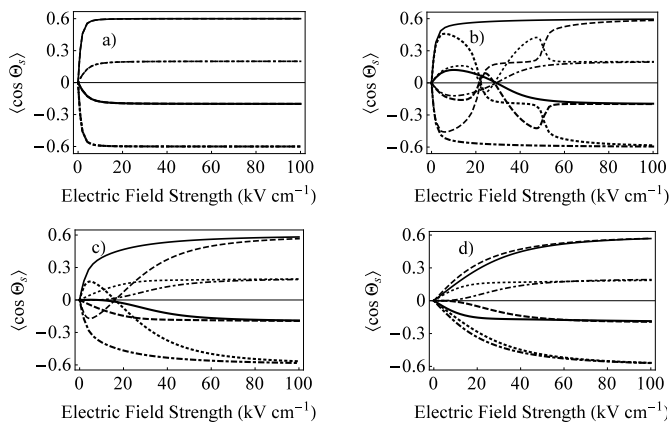}
\caption{$\langle \cos{\Theta} \rangle$ for all eight OH states. The parameters are
$B = 1 T$ and a) $\theta =0$ b) $\theta =\pi/6$ c) $\theta = \pi/3$ and d) $\theta=\pi/2$.
These plots of our analytical results can be compared to Fig. 10 (a)-(d) of Ref.\cite{Sharma2015} which
displays the same quantities obtained by solving Schrodinger's equation with a large basis set. As can
be verified from such a comparison, the analytical agree well with the full numerical calculations.}
\label{fig:F6}
\end{figure}

We first present the orientation for all eight OH ground states, as a function of the applied electric field,
with a constant magnetic field and for various values of the angle of separation. These values have been shown
in Fig.~\ref{fig:F6} (a)-(d)
and may be compared to Fig. 10 (a)-(d) of Ref.~\cite{Sharma2015} which displays the same quantities obtained by
numerically solving Schrodinger's equation with a large number of basis functions. As can be seen from such a
comparison, the analytical results agree well with the full numerical calculations. We note that in this
specific case, to enable comparison to the numerical results, we calculated the molecular axis orientation
with respect to the electric field, rather than with respect to the laboratory $z$ axis. The two cases
are linked through a straightforward coordinate transformation \cite{Sharma2015}.

\begin{figure}[h]
\centering
\includegraphics[width=0.5\textwidth]{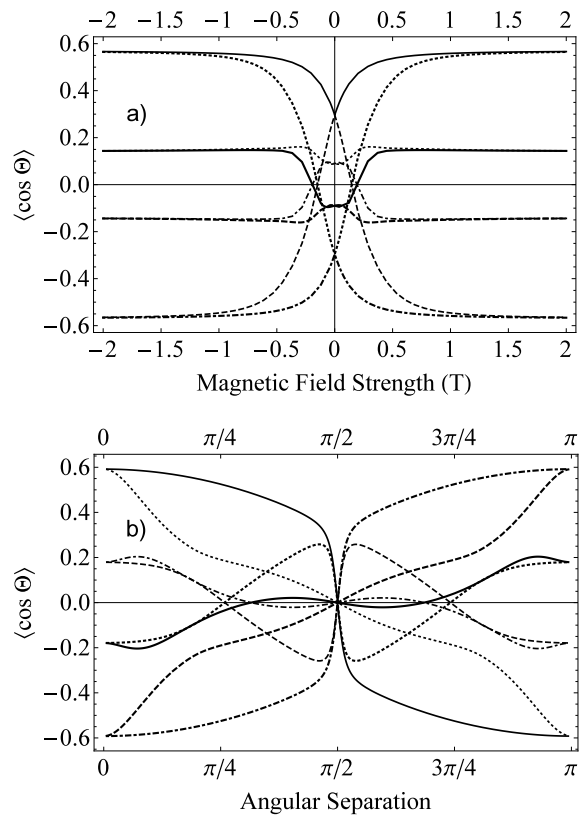}
\caption{$\langle \cos{\Theta} \rangle$ for the 8 OH states. The parameters are a) $\theta =\pi/6$ and
$E = 10 \text{k}V/\text{cm}$ b)  $B = 0.2T$ and $E = 10 \text{k}V/\text{cm}$. Note that the orientation
vanishes at $\theta=\pi/2$ for all states [Eq.~(\ref{eq:CosPerp})].}
\label{fig:F7}
\end{figure}

\begin{figure}[h]
\centering
\includegraphics[width=0.5\textwidth]{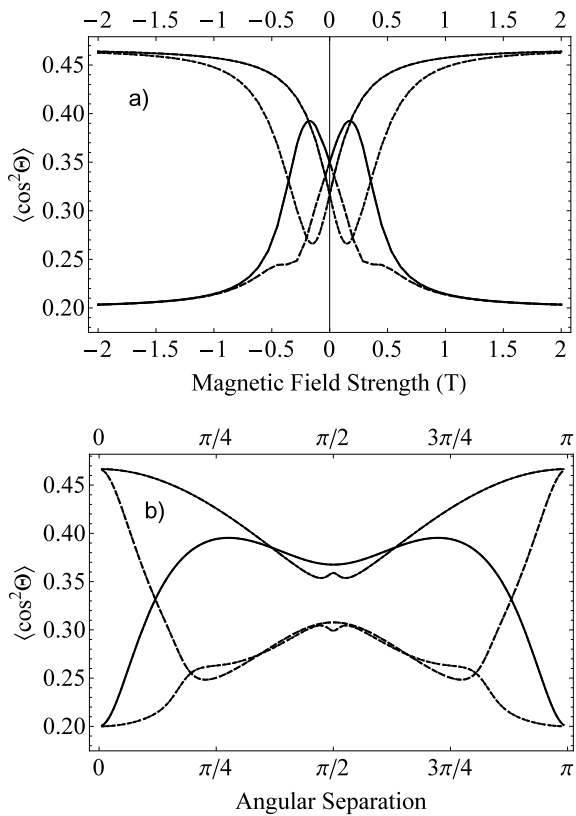}
\caption{$\langle \cos^2{\Theta} \rangle$ for the 8 OH states. The parameters are a) $\theta =\pi/6$ and
$E = 10 \text{k}V/\text{cm}$ b)  $B = 0.2T$ and $E = 10 \text{k}V/\text{cm}$.}
\label{fig:F8}
\end{figure}

Useful analytical approximations can be found even in this most general case. For example, the orientation
of the least energetic state goes linearly with the electric field for small fields, and is to a good
approximation given by
\begin{equation}
\langle \cos{\Theta} \rangle^{(1)} = \frac{18 \cos{\theta}\tilde{E}}{25 \tilde{\Delta}} + \mathcal{O} (\tilde{E^2}).
\end{equation}
We note that this result is independent of the magnetic field; however outside of the linear regime this is no
longer true, and the magnetic field must be taken into consideration. The variation of orientation and
alignment with the magnetic field has likewise been displayed in Figs.~\ref{fig:F7}(a) and ~\ref{fig:F8}(a),
respectively. To our knowledge the variation of OH stereochemical properties with the inter-field angle $\theta$,
has not been presented earlier in the literature. These have been shown in Figs.~\ref{fig:F7}(b) and
~\ref{fig:F8}(b), respectively.
We note that the stereochemical properties display various kinds of degeneracies. For example, whereas
in the most general case, as can be seen from Figs.~\ref{fig:F7} and ~\ref{fig:F8}, the orientation
$\langle \cos{\Theta}\rangle$ is distinct for each state, the alignment $\langle \cos^2{\Theta} \rangle$
is always doubly degenerate. Furthermore we have found (not shown in the plots), that in the presence of
only an electric or magnetic field, or for parallel fields, $\langle \cos{\Theta} \rangle$ is two-fold
degenerate but $\langle \cos^2{\Theta} \rangle$ is four-fold degenerate, i.e. with only two distinct
values. Thus, as the symmetries of the problem are restored, more degeneracy appears in the stereochemical
properties.

\subsection{Higher order moments}
\label{subsec:Higher}
The alignment and orientation are often adequate for describing the angular distribution of the
molecular axis. For some applications, however, a higher level of characterization is needed
\cite{Hockett2015}. In this section we will present the quantities $\langle \cos ^3 \Theta \rangle$
and $\langle \cos ^4 \Theta \rangle$ calculated using our analytical method. Our task is simplified
by the fact that the structure of the matrix representation of these higher order terms is similar
to that of the orientation $(k=1)$ and alignment $(k=2)$ already considered above.

\begin{figure}[h]
\centering
\includegraphics[width=0.5\textwidth]{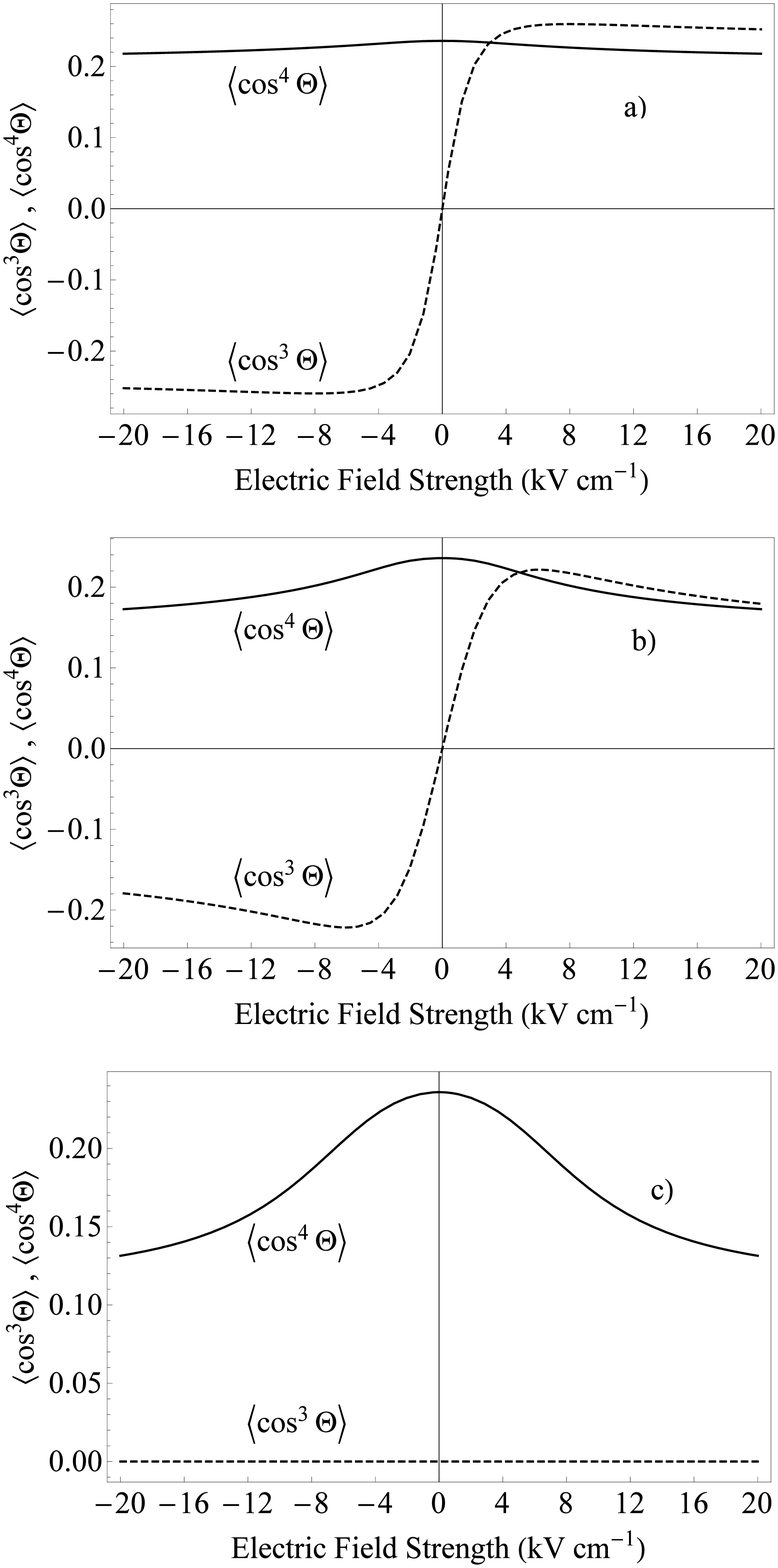}
\caption{$\langle \cos^3{\Theta} \rangle$ (dashed), $\langle \cos^4{\Theta} \rangle$ (solid).
The parameters are $B= 0.2 T$ and a) $\theta =  \pi/6 $ b) $\theta =  \pi/3$ c) $\theta =  \pi/2$.}
\label{fig:F9}
\end{figure}

Let us first examine the case of parallel fields. In this case
\begin{equation}
\langle \cos ^3 \Theta \rangle = \frac{42 \tilde{E}}{25 \sqrt{25 \tilde{\Delta} ^2+36 \tilde{E}^2}},
\end{equation}
and
\begin{equation}
\langle \cos ^4 \Theta \rangle = \frac{289}{1225}.
\end{equation}
As expected, these results are quite similar to their lower order counterparts. However we do notice
that they are scaled down by a factor of approximately one half.
Now let us consider the situation with both fields turned on with an arbitrary angle of separation.
In this case, the expressions are quite long, and examples of the results are presented graphically
in Fig.~\ref{fig:F9}. From these plots, we find, as for the lower order cosines that the degree of
manipulation is maximized in the parallel (perpendicular) configuration for
$\langle \cos^3{\Theta} \rangle$ ($\langle \cos^4{\Theta} \rangle$).

\section{Conclusions}
\label{sec:Con}
From the analytic treatment presented in our article, we arrive at the following conclusions regarding
the stereochemical properties of the ground state OH molecule in crossed electric and magnetic fields:
\begin{enumerate}
\item Analytical calculations based on an eight-dimensional effective matrix Hamiltonian compare well
to full numerical calculations (Section~\ref{subsec:Crossed}).
\item A magnetic field does not, by itself, allow manipulation of the stereochemical properties of OH
(Section ~\ref{subsec:BField}) \cite{Schmelcher2013}. Likewise, if a magnetic field is applied parallel
to an electric field, only the electric field has any stereochemical effect on the molecule
(Section ~\ref{subsec:Parallel}).
\item If it is desirable to manipulate the orientation - but not the alignment - of the molecule, solely
an electric field should be used. In that case a compact expression for the orientation is available
(Section ~\ref{subsec:EField}).
\item If it is desirable to create alignment - but completely cancel the orientation - of the molecule,
one way is to use perpendicular electric and magnetic fields. In that case a reasonably compact expression
for the alignment is available, and yields simple and useful approximations (Section ~\ref{subsec:Perp}).
\item Higher order moments of the molecular axis distribution can readily be found and analyzed using
our analytical approach (Section ~\ref{subsec:Higher}).
\end{enumerate}
We expect these conclusions to be useful guiding principles, even in considerations of fully numerical
work. Future work will investigate the possibilities added by the use of time-varying electromagnetic
fields \cite{Sharma2015} and also probe the connection between orientation-to-alignment conversion and
spin squeezing \cite{Budker2012}.

\section{Acknowledgements}
We are grateful to the JILA groups for past discussions on the OH molecule. S. M. would like to thank
the Rochester Institute of Technology for a College of Science Summer Research Fellowship.

\appendix

\section{Components of the eigenstate}

In this Appendix we supply the expressions for the components of the eigenstate of Eq.~(\ref{eq:eigenv}). In the
expressions below, $\lambda$ is the eigenvalue of the state \cite{Mishkat2013}

\begin{small}
\begin{equation}
\begin{aligned}
c_1 & = \left[2304 \tilde{B}^4+3840 \tilde{B}^3 \tilde{\Delta} +32 \tilde{B}^2 \left(72 \tilde{E}^2-25 \left(\tilde{\Delta} ^2+20 \lambda ^2\right)\right) \right.\\ &
+1152 \tilde{B}^2 \tilde{E}^2 \cos 2\theta -80 \tilde{B} \tilde{\Delta}  \left(25 \left(\tilde{\Delta} ^2-4 \lambda ^2\right)+36 \tilde{E}^2\right)\\
&\left. +\left(25 \left(\tilde{\Delta} ^2-4 \lambda ^2\right)+4 \tilde{E}^2\right) \left(25 \left(\tilde{\Delta} ^2-4 \lambda ^2\right)+36 \tilde{E}^2\right)\right]\\
& \times \left(-48 \tilde{E}^3 \sin ^3\theta \right) ,
\end{aligned}
\end{equation}

\begin{equation}
\begin{aligned}
c_2 &= \left[-20736 \tilde{B}^4+23040 \tilde{B}^3 (\tilde{\Delta} -\lambda )-2400 \tilde{B}^2 \left(3 \tilde{\Delta} ^2-16 \tilde{\Delta}  \lambda \right. \right.  \\
&\left.-20 \lambda ^2\right) +3456 \tilde{B}^2 \tilde{E}^2 \cos 2 \theta-160 \tilde{B} \left(25 (\tilde{\Delta} +2 \lambda ) (\tilde{\Delta} +5 \lambda )  \right. \\
&\times (\tilde{\Delta} -2 \lambda )\left. +36 \tilde{E}^2 (\tilde{\Delta} +\lambda )\right)+432 \tilde{E}^4 \\
& \left. +375 \left(\tilde{\Delta} ^2-4 \lambda ^2\right) \left(5 \left(\tilde{\Delta} ^2-4 \lambda ^2\right)+8 \tilde{E}^2\right) \right]\\
&\times \left(-16 \sqrt{3} \tilde{E}^3 \sin ^2\theta  \cos \theta\right),
\end{aligned}
\end{equation}

\begin{equation}
\begin{aligned}
c_3  & =\left[ -3456 \tilde{B}^2 \tilde{E}^4 \sin 5 \theta+2 \tilde{E}^2 \sin 3 \theta  \left(6912 \tilde{B}^4-23040 \tilde{B}^3 \tilde{\Delta} \right.\right. \\
& \left.  +192 \tilde{B}^2 \left(25 (3 \tilde{\Delta} +4 \lambda ) (\tilde{\Delta} -4 \lambda )+9 \tilde{E}^2\right)+80 \tilde{B} \left(25 (\tilde{\Delta} +2 \lambda ) \right. \right.  \\
& \left. \left. (\tilde{\Delta} +20 \lambda ) (\tilde{\Delta} -2 \lambda )+36 \tilde{E}^2 (\tilde{\Delta} +4 \lambda )\right)-3 \left(25 \left(\tilde{\Delta} ^2-4 \lambda ^2\right) \right. \right. \\
& \left. \left.  +4 \tilde{E}^2\right) \left(25 \left(\tilde{\Delta} ^2-4 \lambda ^2\right)+36 \tilde{E}^2\right)\right)+\sin \theta \left(-110592 \tilde{B}^6 \right.  \\
& \left. -92160 \tilde{B}^5 (\tilde{\Delta} +4 \lambda )-768 \tilde{B}^4 \left(-325 \tilde{\Delta} ^2+800 \tilde{\Delta}  \lambda +18 \tilde{E}^2 \right. \right.  \\
 & \left. \left. -700 \lambda ^2\right)+1280 \tilde{B}^3 \left(25 (\tilde{\Delta} +4 \lambda ) \left(5 \tilde{\Delta} ^2-16 \tilde{\Delta}  \lambda +20 \lambda ^2\right) \right. \right.   \\
 &  \left. \left. +144 \tilde{E}^2 (\tilde{\Delta} -\lambda )\right)+16 \tilde{B}^2 \left(-625 (\tilde{\Delta} -2 \lambda ) (\tilde{\Delta} +2 \lambda ) \left(13 \tilde{\Delta} ^2 \right. \right. \right. \\
 & \left. \left. \left. -32 \tilde{\Delta}  \lambda +28 \lambda ^2\right)+864 \tilde{E}^4-2400 \tilde{E}^2 (3 \tilde{\Delta} -7 \lambda ) (\tilde{\Delta} -\lambda )\right) \right. \\
 &\left.  -40 \tilde{B} \left(625 (\tilde{\Delta} +4 \lambda ) \left(\tilde{\Delta} ^2-4 \lambda ^2\right)^2+576 \tilde{\Delta}  \tilde{E}^4+100 \tilde{E}^2 \right. \right.  \\
 & \left. \left.  (\tilde{\Delta} +2 \lambda ) (13 \tilde{\Delta} +20 \lambda ) (\tilde{\Delta} -2 \lambda )\right)+\left(25 \left(\tilde{\Delta} ^2-4 \lambda ^2\right)+4 \tilde{E}^2\right) \right. \\
 &\left. \times \left.  \left(25 \left(\tilde{\Delta} ^2-4 \lambda ^2\right)+6 \tilde{E}^2\right) \left(25 \left(\tilde{\Delta} ^2-4 \lambda ^2\right)+36 \tilde{E}^2\right)\right)\right]\\
 &\times\left( 2 \sqrt{3} \tilde{E}\right),
\end{aligned}
\end{equation}

\begin{equation}
\begin{aligned}
c_4 & = \left[  -1152 \tilde{B}^2 \tilde{E}^4 \cos 5\theta+2 \tilde{E}^2 \cos 3\theta \left(11520 \tilde{B}^4+30720 \tilde{B}^3 \lambda \right. \right.\\
& -64 \tilde{B}^2 \left(25 \left(8 \lambda ^2-5 \tilde{\Delta} ^2\right)+9 \tilde{E}^2\right)+160 \tilde{B} \lambda  \left(125 \left(\tilde{\Delta} ^2-4 \lambda ^2\right) \right. \\
& \left. \left.  +36 \tilde{E}^2\right) -\left(25 \left(\tilde{\Delta} ^2-4 \lambda ^2\right)+4 \tilde{E}^2\right) \left(25 \left(\tilde{\Delta} ^2-4 \lambda ^2\right)+36 \tilde{E}^2\right)\right) \\
& +\cos \theta \left(-36864 \tilde{B}^6-61440 \tilde{B}^5 \lambda +256 \tilde{B}^4 \left(475 \tilde{\Delta} ^2+18 \tilde{E}^2 \right. \right. \\
& \left. \left. +1700 \lambda ^2\right)+7680 \tilde{B}^3 \lambda  \left(25 \left(\tilde{\Delta} ^2+4 \lambda ^2\right)-4 \tilde{E}^2\right)-16 \tilde{B}^2  \left(625 \right. \right.   \\
& \times \left. \left(11 \tilde{\Delta} ^4  -72 \tilde{\Delta} ^2 \lambda ^2+112 \lambda ^4\right)+288 \tilde{E}^4+2400 \tilde{E}^2 \left(2 \tilde{\Delta} ^2-\lambda ^2\right)\right)\\
&  -80 \tilde{B} \lambda  \left(1875 \left(\tilde{\Delta} ^2-4 \lambda ^2\right)^2+192 \tilde{E}^4+1100 \tilde{E}^2 \left(\tilde{\Delta} ^2-4 \lambda ^2\right)\right)\\
& +\left(25 \left(\tilde{\Delta} ^2-4 \lambda ^2\right)+4 \tilde{E}^2\right) \left(25 \left(\tilde{\Delta} ^2-4 \lambda ^2\right)+6 \tilde{E}^2\right) \\
&\left.\left.  \times \left(25 \left(\tilde{\Delta} ^2-4 \lambda ^2\right)+36 \tilde{E}^2\right)\right)\right]\times \left(6 \tilde{E} \right),
\end{aligned}
\end{equation}

\begin{equation}
\begin{aligned}
c_5 &= \left[144 \tilde{B}^2+25 (\tilde{\Delta} +2 \lambda ) (\tilde{\Delta} -10 \lambda )+36 \tilde{E}^2\right]\\
&\times\left(-1536 \tilde{B} \tilde{E}^4 \sin ^3\theta  \cos \theta\right),
\end{aligned}
\end{equation}

\begin{equation}
\begin{aligned}
c_6  &=\left[27648 \tilde{B}^5+11520 \tilde{B}^4 (5 \tilde{\Delta} +2 \lambda )+384 \tilde{B}^3 \left(25 \left(\tilde{\Delta} ^2+4 \tilde{\Delta}  \lambda  \right. \right.\right.  \\
& \left. \left. -20 \lambda ^2\right)  +36 \tilde{E}^2\right)-160 \tilde{B}^2 \left(25 (\tilde{\Delta} +2 \lambda ) \left(7 \tilde{\Delta} ^2-12 \tilde{\Delta}  \lambda +20 \lambda ^2\right) \right. \\
& \left. +144 \tilde{E}^2 (\tilde{\Delta} -\lambda )\right)-4 \tilde{B} \left(25 (\tilde{\Delta} +2 \lambda ) (\tilde{\Delta} +6 \lambda )+36 \tilde{E}^2\right) \\
& \left(25 \left(\tilde{\Delta} ^2-4 \lambda ^2\right)+12 \tilde{E}^2\right)+96 \tilde{B} \tilde{E}^2 \cos 2 \theta  \left(5 (\tilde{\Delta} +2 \lambda ) (12 \tilde{B} \right. \\
& \left. -5 \tilde{\Delta} +50 \lambda )-36 \tilde{E}^2\right)+5 (\tilde{\Delta} +2 \lambda ) \left(25 \left(\tilde{\Delta} ^2-4 \lambda ^2\right)+4 \tilde{E}^2\right) \\
& \left.\times \left(25 \left(\tilde{\Delta} ^2-4 \lambda ^2\right)+36 \tilde{E}^2\right)\right] \times \left(8 \sqrt{3} \tilde{E}^2 \sin ^2\theta \right),
\end{aligned}
\end{equation}

\begin{equation}
\begin{aligned}
c_7 & =  \left[-13824 \tilde{B}^5+11520 \tilde{B}^4 (\tilde{\Delta} -6 \lambda )+192 \tilde{B}^3 \left(25 \left(5 \tilde{\Delta} ^2 \right. \right. \right. \\
& \left. \left. \left. +16 \tilde{\Delta}  \lambda -12 \lambda ^2\right)+36 \tilde{E}^2\right)-160 \tilde{B}^2 \left(25 (\tilde{\Delta} +2 \lambda ) \left(5 \tilde{\Delta} ^2- \right. \right. \right. \\
& \left. \left. \left. 12 \tilde{\Delta}  \lambda -20 \lambda ^2\right)+72 \tilde{\Delta}  \tilde{E}^2\right)+2 \tilde{B} \left(-625 (\tilde{\Delta} -2 \lambda ) (\tilde{\Delta} +2 \lambda )^2 \right. \right.  \\
&\times \left. \left. (3 \tilde{\Delta} +26 \lambda )+432 \tilde{E}^4-2400 \tilde{E}^2 (\tilde{\Delta} +2 \lambda ) (\tilde{\Delta} +4 \lambda )\right) \right. \\
& \left.   +5 (\tilde{\Delta} +2 \lambda ) \left(25 \left(\tilde{\Delta} ^2-4 \lambda ^2\right)+4 \tilde{E}^2\right) \left(25 \left(\tilde{\Delta} ^2-4 \lambda ^2\right)+36 \tilde{E}^2\right)\right)\\
& +24 \tilde{B} \tilde{E}^2 \sin 4\theta \left(5 (\tilde{\Delta} +2 \lambda ) (24 \tilde{B}-5 \tilde{\Delta} +50 \lambda )-36 \tilde{E}^2\right]\\
&\left(8 \sqrt{3} \tilde{E}^2 \sin 2 \theta \right),
\end{aligned}
\end{equation}

\begin{equation}
\begin{aligned}
c_8 & =  -\left[442368 \tilde{B}^7-184320 \tilde{B}^6 (\tilde{\Delta} -2 \lambda )+3072 \tilde{B}^5 \left(36 \tilde{E}^2 \right. \right. \\
& \left. \left. -25 \left(19 \tilde{\Delta} ^2+4 \tilde{\Delta}  \lambda +76 \lambda ^2\right)\right)+1280 \tilde{B}^4 \left(25 (\tilde{\Delta} -2 \lambda ) \right. \right.  \\
&\left. \left.  \left(19 \tilde{\Delta} ^2+4 \tilde{\Delta}  \lambda +76 \lambda ^2\right)+288 \tilde{E}^2 \lambda \right)+192 \tilde{B}^3 \left(625 (\tilde{\Delta} +2 \lambda )^2 \right. \right.  \\
&\times \left. \left.  \left(11 \tilde{\Delta} ^2-36 \tilde{\Delta}  \lambda +44 \lambda ^2\right)+576 \tilde{E}^4+400 \tilde{E}^2 \left(17 \tilde{\Delta} ^2 \right. \right. \right.  \\
& \left. \left. \left. +2 \tilde{\Delta}  \lambda -28 \lambda ^2\right)\right)-80 \tilde{B}^2 \left(625 (\tilde{\Delta} +2 \lambda )^2 (\tilde{\Delta} -2 \lambda ) \left(11 \tilde{\Delta} ^2   \right. \right. \right.\\
& \left. \left. \left. -36 \tilde{\Delta}  \lambda +44 \lambda ^2\right)+576 \tilde{E}^4 (3 \tilde{\Delta} -5 \lambda )+600 \tilde{E}^2 (\tilde{\Delta} +2 \lambda ) \right. \right.  \\
& \times \left. \left.\left(13 \tilde{\Delta} ^2-40 \tilde{\Delta}  \lambda +44 \lambda ^2\right)\right)+192 \tilde{B} \tilde{E}^4 \cos 4\theta \left(144 \tilde{B}^2 \right. \right.  \\
&  \left. \left. -180 \tilde{B} (\tilde{\Delta} +2 \lambda )+25 (\tilde{\Delta} +2 \lambda ) (\tilde{\Delta} -10 \lambda )+36 \tilde{E}^2\right) \right.  \\
&\left.  -4 \tilde{E}^2 \cos 2\theta  \left(110592 \tilde{B}^5-11520 \tilde{B}^4 (3 \tilde{\Delta} -10 \lambda ) \right. \right.  \\
& \left. \left.  +1152 \tilde{B}^3 \left(125 \left(\tilde{\Delta} ^2-4 \lambda ^2\right)+12 \tilde{E}^2\right)-640 \tilde{B}^2 \left(25 (\tilde{\Delta} -\lambda ) \right. \right. \right. \\
& \times \left. \left. \left. (\tilde{\Delta} +2 \lambda ) (5 \tilde{\Delta} -16 \lambda )+18 \tilde{E}^2 (3 \tilde{\Delta} -4 \lambda )\right)+24 \tilde{B} \right. \right.  \\
& \left. \left. \times \left(-625 (\tilde{\Delta} -2 \lambda ) (\tilde{\Delta} +2 \lambda )^2 (\tilde{\Delta} +10 \lambda )+144 \tilde{E}^4 \right. \right. \right.  \\
& \left. \left. \left.  -400 \tilde{E}^2 (\tilde{\Delta} +2 \lambda ) (2 \tilde{\Delta} +7 \lambda )\right)+15 (\tilde{\Delta} +2 \lambda ) \left(25 \left(\tilde{\Delta} ^2 \right. \right. \right. \right.  \\
& \left. \left. \left. \left. -4 \lambda ^2\right)+4 \tilde{E}^2\right) \left(25 \left(\tilde{\Delta} ^2-4 \lambda ^2\right)+36 \tilde{E}^2\right)\right)-300 \tilde{B} (\tilde{\Delta} \right. \\
& \left. +2 \lambda ) \left(625 (\tilde{\Delta} +2 \lambda )^3 (\tilde{\Delta} -2 \lambda )^2+192 \tilde{E}^4 (3 \tilde{\Delta} +\lambda ) \right. \right. \\
&\left. \left.  +100 \tilde{E}^2 (\tilde{\Delta} +2 \lambda ) (13 \tilde{\Delta} +6 \lambda ) (\tilde{\Delta} -2 \lambda )\right)+5 (\tilde{\Delta} +2 \lambda )\right.  \\
& \left. \times \left(25 \left(\tilde{\Delta} ^2-4 \lambda ^2\right)+4 \tilde{E}^2\right) \left(25 \left(\tilde{\Delta} ^2-4 \lambda ^2\right)+16 \tilde{E}^2\right) \right.  \\
\nonumber
\end{aligned}
\end{equation}

\begin{equation}
\begin{aligned}
& \times \left. \left(25 \left(\tilde{\Delta} ^2-4 \lambda ^2\right)+36 \tilde{E}^2\right)\right].
\end{aligned}
\end{equation}

\end{small}


\end{document}